\documentstyle[12pt,epsf,a4]{article}

\newcommand{\beq}{\begin{equation}}
\newcommand{\eeq}{\end{equation}}
\newcommand{\beqa}{\begin{eqnarray}}
\newcommand{\eeqa}{\end{eqnarray}}
\newcommand{\no}{\nonumber}
\newcommand{\q}{\quad}
\newcommand{\qq}{\qquad}
\newcommand{\mnod}{\stackrel{\circ}{M}}

\begin{document}

\hfill 

\hfill 

\bigskip\bigskip

\begin{center}

{{\Large\bf The electric dipole moment of the neutron in chiral perturbation
theory \footnote{Work
supported in part by the BMBF}}}

\end{center}

\vspace{.4in}

\begin{center}
{\large B. Borasoy\footnote{email: borasoy@physik.tu-muenchen.de}}

\bigskip

\bigskip

Physik Department\\
Technische Universit{\"a}t M{\"u}nchen\\
D-85747 Garching, Germany \\

\vspace{.2in}

\end{center}

\vspace{.7in}

\thispagestyle{empty} 

\begin{abstract}
We calculate the electric dipole moments of the neutron and the $\Lambda$
within the framework of heavy baryon chiral perturbation theory. They are
induced by strong $CP$-violating terms of the effective Lagrangian in the
presence  of the vacuum angle $\theta_0$. The construction of such a Lagrangian
is outlined and we are able to give an estimate for $\theta_0$. 
\end{abstract}

\vfill

\section{Introduction} 
The axial $U(1)$ anomaly in QCD implies an additional term in the Lagrangian,
which violates $P$, $T$ and $CP$. This new term is proportional to the
so-called vacuum angle $\theta_0$, an unknown parameter, and its size may be
determined from $CP$-violating effects, as e.g. $\eta \rightarrow \pi \pi$ or
the electric dipole moments of the neutron and the $\Lambda$. 
The most recent measurements of
the  electric dipole moment of the neutron 
$d_n^\gamma$ have constrained it to \cite{Har}
\beq  \label{exp}
|d_n^\gamma| < 6.3  \times 10^{-26} \, e \, \mbox{cm} \, .
\eeq
On the other hand, theoretical estimates for $d_n^\gamma$ induced by the
$\theta_0$-term can be given leading to an upper bound for $\theta_0$ 
\cite{the,PR,sum}.
Of particular interest here is the estimate of Pich and de Rafael 
\cite{PR} who used an effective chiral Lagrangian approach and came to the
conclusion that it is possible to obtain an estimate for the size of the vacuum
angle $\theta_0$ with an experimental upper limit of 
$| \theta_0 | \le 5 \times 10^{-10}$. However, the authors worked in a
relativistic framework which does not have a systematic chiral counting scheme,
so that higher loop diagrams contribute to lower chiral orders. This problem is
avoided in heavy baryon chiral perturbation theory as proposed in \cite{JM},
which allows for a consistent power counting. One should therefore study this
effect within the heavy baryon formulation.
Furthermore, the baryon  Lagrangian in \cite{PR} 
which describes the interactions of
the neutron with the pseudoscalar nonet ($\pi, K, \eta, \eta'$) 
does not contain
explicitly $CP$-violating terms. They are rather induced by the vacuum
alignment of the purely mesonic Lagrangian in the presence of the
$\theta_0$-term. 
As shown in \cite{B1, B2} the most general baryonic Lagrangian taking the
axial $U(1)$ anomaly into account does have explicitly $CP$-violating terms
even at lower chiral orders. It has to be checked, if such terms lead to
sizeable contributions for the present consideration.
Finally, the authors of \cite{PR} proposed to estimate the contribution from
unknown counter\-terms by varying the scale in the chiral logarithms. This
procedure reveals the scale dependence of the involved coupling constants but
not their absolute value, and it is desirable to have a somewhat more reliable
estimate of the involved couplings. 
The aim of the present work is to reinvestigate the electric dipole moments of
the neutron and the $\Lambda$
by taking the above mentioned points into consideration. One 
has to check, if it is still possible to give a reliable estimate of the vacuum
angle $\theta_0$, and if it is different from the one given in \cite{PR}.

The paper is organized as follows. In the next section we present the purely
mesonic effective Lagrangian in the presence of the $\theta_0$-term and the
vacuum alignment is discussed. Baryon fields are included in the effective
theory in Sec. 3 by using the method outlined in \cite{B1}. We proceed by
calculating in Sec. 4 the electric dipole moment of the neutron 
and the $\Lambda$ up to one-loop
order within the heavy baryon framework.
Numerical results and conclusions are given in Sec. 5.

\section{The mesonic Lagrangian}
In this section, we will consider the purely mesonic Lagrangian in the presence
of the $\theta_0$-term.
The derivation of this Lagrangian has been given elsewhere, see e.g.
\cite{GL,Leu,H-S}, so we will restrict ourselves to the 
repetition of some of the
basic formulae which are needed in the present work. In \cite{Leu,H-S}
the topological charge operator coupled to an external field is added to the
QCD Lagrangian
\beq  \label{lag}
{\cal L} = {\cal L}_{QCD} - \frac{g^2}{16 \pi^2} \theta(x) \mbox{tr}_c
  ( G_{\mu \nu} \tilde{G}^{\mu \nu} )
\eeq
with $\tilde{G}_{\mu \nu} = \epsilon_{\mu \nu \alpha \beta} G^{\alpha
\beta}$ and $\mbox{tr}_c$ is the trace over the color indices.
Under $U(1)_R \times U(1)_L$ the axial $U(1)$ anomaly 
adds a term $ -( g^2 / 16 \pi^2)
2 N_f \, \alpha \, \mbox{tr}_c ( G_{\mu \nu} \tilde{G}^{\mu \nu} )$ to the
QCD Lagrangian, with $N_f$ being the number of different quark flavors and
$\alpha$ the angle of the global axial $U(1)$ rotation.
The vacuum angle $\theta(x)$ is in this context treated as an external field
that transforms under an axial $U(1)$ rotation as
\beq
\theta(x) \rightarrow  \theta'(x) = \theta(x) - 2 N_f \alpha .
\eeq
Then the term generated by the anomaly in the fermion determinant is
compensated by the shift in the $\theta$ source and the Lagrangian from
Eq. (\ref{lag}) remains invariant under axial $U(1)$ transformations.
The symmetry group $SU(3)_R \times SU(3)_L$ of the Lagrangian ${\cal L}_{QCD}$
is extended\footnote{To be more
precise, the Lagrangian changes by a total derivative which gives rise to the
Wess-Zumino term. We will neglect this contribution since the corresponding
terms involve five or more meson fields which do not play any role for the
discussions here.} to $U(3)_R \times U(3)_L$ for ${\cal L}$.
The Green functions of QCD are obtained by expanding the generating functional
around $\theta(x) = \theta_0$ where the phase of the quark mass matrix emerging
from the Yukawa couplings of the light quarks in the electroweak sector has
been absorbed in $\theta_0$.
The extended symmetry remains at the level of an effective theory and the
additional source $\theta$ also shows up in the effective Lagrangian.
Let us consider the purely mesonic effective theory first.
The lowest lying pseudoscalar meson nonet is summarized in a matrix valued 
field $\tilde{U}(x)$.
The effective Lagrangian is formed with the fields $\tilde{U}(x)$,  
derivatives thereof and also includes both the quark mass matrix ${\cal M}$ and
the vacuum angle $\theta$: ${\cal L}_{\mbox{eff}}(\tilde{U},\partial 
\tilde{U},\ldots,
{\cal M},\theta)$. Under $U(3)_R \times U(3)_L$ the fields transform as
follows:
\beq
\tilde{U}' = R\tilde{U}L^{\dagger} \q , 
\qq {\cal M}'= R{\cal M}L^{\dagger} \q , \qq
\theta'(x) = \theta(x) - 2 N_f \alpha
\eeq
with $R \in U(3)_R$, $L \in U(3)_L$,
but the Lagrangian remains invariant. 
The phase of the determinant of $\tilde{U}(x)$
transforms  under axial $U(1)$ as
$ \ln \mbox{det} \tilde{U}'(x)=  \ln \mbox{det} \tilde{U}(x) + 
2 i N_f \alpha$ so that
the combination $\theta -i \ln \mbox{det} \tilde{U}$ remains invariant.
It is more convenient to replace the variable $\theta$ by this invariant
combination, ${\cal L}_{\mbox{eff}}(\tilde{U},\partial \tilde{U},\ldots,
{\cal M},\theta -i \ln \mbox{det} \tilde{U})$.
One can now construct the effective Lagrangian in these
fields that respects the symmetries of the underlying theory.
In particular, the Lagrangian is invariant under $U(3)_R \times U(3)_L$
rotations of $\tilde{U}$ and ${\cal M}$ at a fixed value of the last argument.
The Lagrangian up to and including terms with two derivatives and
one factor of ${\cal M}$ reads
\beqa  \label{mes}
{\cal L}_{\phi} &=& - V_0 + V_1 \langle \nabla_{\mu}\tilde{U}^{\dagger} 
\nabla^{\mu} \tilde{U} 
\rangle  + V_2 \langle \tilde{\chi}^{\dagger} \tilde{U} + 
 \tilde{\chi} \tilde{U}^{\dagger}  \rangle 
+ i V_3 \langle \tilde{\chi}^{\dagger} \tilde{U} - 
  \tilde{\chi} \tilde{U}^{\dagger}  \rangle \no \\
&&+ V_4 \langle \tilde{U} \nabla_{\mu}\tilde{U}^{\dagger}  \rangle 
\langle \tilde{U}^{\dagger} \nabla^{\mu}\tilde{U} \rangle .
\eeqa
The expression $\langle \ldots \rangle$ denotes the trace in flavor space
and $\tilde{\chi}$ is proportional to the quark mass matrix 
$\tilde{\chi} = \tilde{\chi}^{\dagger} = 2 B_0 {\cal M}$ 
with ${\cal M} = \mbox{diag}(m_u,m_d,m_s)$
and $B_0 = - \langle  0 | \bar{q} q | 0\rangle/ F_\pi^2$ the order
parameter of the spontaneous symmetry violation.
The covariant derivative of $\tilde{U}$ is defined by
\beq
\nabla_{\mu} \tilde{U}  =  \partial_{\mu} \tilde{U} 
                     - i ( v_{\mu} + a_{\mu}) \tilde{U}
                     + i \tilde{U} ( v_{\mu} - a_{\mu})  . 
\eeq
The external fields $v_{\mu}(x),a_{\mu}(x)$ represent Hermitian $3 \times 3$
matrices in flavor space.
Note that a term of the type $i V_5\langle \tilde{U}^{\dagger} \nabla_{\mu}
\tilde{U} \rangle \nabla^{\mu} \theta$
can be transformed away \cite{Leu} 
and a term proportional to $V_6 \nabla_{\mu} \theta \nabla^{\mu} \theta$ 
does not enter the calculations performed in the present
work and will be neglected.
The coefficients $V_i$ are functions of the variable 
$\theta -i \ln \mbox{det} \tilde{U}$, $V_i(\theta -i \ln \mbox{det} 
\tilde{U})$,
and can be expanded in terms of this variable. 
The terms $V_{1,\ldots,4}$ are of second chiral order, whereas $V_0$
is of zeroth chiral order.
Parity conservation implies that the $V_i$ are all even functions
of $\theta -i \ln \mbox{det} \tilde{U}$ except $V_3$, which is odd, and
$V_1(0) = V_2(0) = F_\pi^2/4$ gives the correct  normalizaton
for the quadratic terms of the Goldstone boson octet,
where $F_\pi \simeq 92.4$ MeV is the pion decay constant.

In order to use the effective Lagrangian, one must first determine the vacuum
expectation value of $ \tilde{U}$ by minimizing the potential energy
\beq
V(\tilde{U}) =  V_0 - V_2 \langle 
  \tilde{\chi}^{\dagger} \tilde{U} + \tilde{\chi} \tilde{U}^{\dagger}  \rangle 
   - i V_3 \langle \tilde{\chi}^{\dagger}  \tilde{U} - 
         \tilde{\chi} \tilde{U}^{\dagger}  \rangle .
\eeq
Since $\tilde{\chi}$ is 
diagonal, one can assume the minimum $U_0$ to be diagonal as
well and of the form
\beq
U_0 = \mbox{diag}( \mbox{e}^{-i \varphi_u}, \mbox{e}^{-i \varphi_d},
    \mbox{e}^{-i \varphi_s} ) .
\eeq
In terms of the angles $\varphi_q$ the potential becomes
\beq
V(U_0) =  V_0(\bar{\theta}_0) 
           - 4 \, B_0 \, V_2(\bar{\theta}_0) \sum_q m_q \cos \varphi_q
           - 4 \, B_0 \, V_3(\bar{\theta}_0) \sum_q m_q \sin \varphi_q  ,
\eeq
where we have introduced the notation $ \bar{\theta}_0 = \theta_0 
- \sum_q \varphi_q$.
The Taylor expansions of the functions $V_i$ read
\beqa
V_i(\bar{\theta}_0) & = & \sum_{n=0}^\infty V_i^{(2n)} \bar{\theta}_0^{2n} 
\qq \q
\mbox{for} \; i = 0,2  \no \\
V_3(\bar{\theta}_0) & = & \sum_{n=0}^\infty V_i^{(2n+1)} \bar{\theta}_0^{2n+1}
\eeqa
with coefficients not fixed by chiral symmetry. Minimizing the potential with
respect to the angles $\varphi_q$ leads to
\beq  \label{cof}
2 B_0 \, m_q \, \sin \varphi_q = {\cal A} + 2 B_0  \, {\cal B} \, m_q 
 \, \cos \varphi_q
\eeq
with 
\beqa 
{\cal A} & = &  2B_0 \Big( \sum_{n=0}^\infty V_2^{(2n)}\,  \bar{\theta}_0^{2n} 
               \Big)^{-1} \q  \sum_{n=1}^\infty
             \Big( \frac{1}{ 2 B_0} n \,  V_0^{(2n)}\, 
               \bar{\theta}_0^{2n-1}  \no \\
  & & -  2 n \, V_2^{(2n)} \, \bar{\theta}_0^{2n-1}
       \sum_{j=u,d,s} m_j \cos \varphi_j \no \\
  & & -   ( 2n -1) V_3^{(2n-1)} \bar{\theta}_0^{2n-2}
       \sum_{j=u,d,s} m_j \sin \varphi_j \Big)
\eeqa
and
\beq
{\cal B} = \Big( \sum_{n=0}^\infty V_2^{(2n)} \, \bar{\theta}_0^{2n} \Big)^{-1}
            \sum_{n=0}^\infty V_3^{(2n+1)} \, \bar{\theta}_0^{2n+1} .
\eeq
To lowest order both in the quark masses $m_q$ and 1/$N_c$ Eq. (\ref{cof})
reads
\beq
\frac{1}{2} B_0 \, F_\pi^2\,  m_q \sin \varphi_q = V_0^{(2)}\,  \bar{\theta}_0 
\eeq
which is the equation for the $\varphi_q$ considered in \cite{PR, Ven}.
One then writes
\beq
\tilde{U} = \sqrt{U_0} \, U \, \sqrt{U_0}
\eeq
and $U$ can be parametrized as
\begin{equation}
 U(\phi,\eta_0) = 
\exp \lbrace 2 i \phi / F_\pi + i \sqrt{\frac{2}{3}} \eta_0/ F_0 \rbrace  
, 
\end{equation}
where the singlet
$\eta_0$ couples to the singlet axial current with strength $F_0$.
The unimodular part of the field $U(x)$ contains the degrees of freedom of
the Goldstone boson octet $\phi$
\begin{eqnarray}
 \phi =  \frac{1}{\sqrt{2}}  \left(
\matrix { {1\over \sqrt 2} \pi^0 + {1 \over \sqrt 6} \eta_8
&\pi^+ &K^+ \nonumber \\
\pi^-
        & -{1\over \sqrt 2} \pi^0 + {1 \over \sqrt 6} \eta_8 & K^0
        \nonumber \\
K^-
        &  \bar{K^0}&- {2 \over \sqrt 6} \eta_8  \nonumber \\} 
\!\!\!\!\!\!\!\!\!\!\!\!\!\!\! \right) \, \, \, \, \, ,  
\end{eqnarray}
while the phase det$U(x)=e^{i\sqrt{6}\eta_0/F_0}$
describes the $\eta_0$ .
The diagonal subgroup $U(3)_V$ of $U(3)_R \times U(3)_L$ 
does not have a dimension-nine
irreducible representation and consequently does not exhibit a nonet symmetry. 
We have therefore used the different notation $F_0$ for the decay constant 
of the singlet field.

One can now express the effective Lagrangian in terms of the Goldstone boson
matrix $U$ and the angles $\varphi_q$
\beqa \label{lagran}
{\cal L}_{\phi} &=& - V_0 + V_1 \langle \nabla_{\mu}U^{\dagger} 
\nabla^{\mu} U 
\rangle  + [V_2 +{\cal B} V_3]  \langle \chi (U + U^{\dagger} ) \rangle 
- i {\cal A} V_2 \langle U - U^{\dagger} \rangle \no \\
&+ &  i [ V_3 -  {\cal B} V_2 ] \langle \chi (U - U^{\dagger} ) \rangle 
+ {\cal A} V_3 \langle U + U^{\dagger} \rangle 
+ V_4 \langle U \nabla_{\mu}U^{\dagger}  \rangle 
\langle U^{\dagger} \nabla^{\mu}U \rangle ,
\eeqa
where we have absorbed the angles $\varphi_q$ in hermitian matrices
$\chi$ and $H$ by defining
\beq
\sqrt{U_0^\dagger}\, \tilde{\chi} \, \sqrt{U_0^\dagger} = \chi + i H , \qq \qq 
\sqrt{U_0} \, \tilde{\chi}^\dagger \, \sqrt{U_0} = \chi - i H ,
\eeq
so that $ \chi = 2 B_0 \mbox{diag}( m_q \cos \varphi_q)$ and
$H = 2 B_0 \mbox{diag}( m_q \sin \varphi_q) = {\cal A} + {\cal B} \chi$.
The $V_i$ are functions of $\sqrt{6} \eta_0/F_0 + \bar{\theta}_0$,
$V_i(\sqrt{6} \eta_0/F_0 + \bar{\theta}_0)$ and we have assumed the external
fields $v_\mu$ and $a_\mu$ to be diagonal which is the case if one considers
electromagnetic interactions.
Note that the Goldstone boson masses at lowest chiral order
are not only functions of the current
quark masses $m_q$, but also depend on the angles $\cos \varphi_q$.
The kinetic energy of the $\eta_0$ singlet field obtains contributions
from $V_1 \langle \nabla_{\mu}U^{\dagger} \nabla^{\mu} U \rangle  $ and
$V_4 \langle U \nabla_{\mu}U^{\dagger}  \rangle 
\langle U^{\dagger} \nabla^{\mu}U \rangle$ which read
\beq
\Big( \frac{F_\pi^2}{2 F_0^2} + \frac{6}{F_0^2} V_4(0) \Big) 
\partial_\mu \eta_0 \partial^\mu \eta_0  .
\eeq
We renormalize the $\eta_0$ field in such a way that the coefficient in
brackets is 1/2 in analogy to the kinetic term of the octet.
By redefining $F_0$ and keeping for simplicity the same notation both
for $\eta_0$ and $F_0$ one arrives at the same Lagrangian as in 
Eq. (\ref{lagran}) but with  $V_4(0) = (F_0^2-F_\pi^2)/12$ in order to ensure
the usual normalization for the kinetic term of a pseudoscalar particle.

\section{$CP$-violating terms in the baryon Lagrangian}
As mentioned before, another source for $CP$-violation is the baryon
Lagrangian. The $CP$-violating terms can be divided into two groups. Firstly, 
the vacuum alignment of the mesonic Lagrangian induces $CP$ non-conserving
meson-baryon interactions as considered in \cite{PR}. But secondly, there are
also explicitly $CP$-violating terms in the most general Lagrangian in the
presence of the $\theta$-vacuum angle which have been neglected in \cite{PR}.
In order to construct the Lagrangian in the baryon-sector, one has to adopt a
non-linear representations for baryons. The main ingredient for a non-linear
realization is the compensator field $K(\tilde{U},R,L) \in U(3)_V$, which
appears in the chiral $U(3)_L \times U(3)_R$
transformation of the left and right coset representatives,
$\tilde{\xi}_L(\tilde{U})$ and $\tilde{\xi}_R(\tilde{U})$:
\beqa
\tilde{\xi}_L(\tilde{U})  & \rightarrow & L \, \tilde{\xi}_L(\tilde{U}) \, 
             K^\dagger (\tilde{U},R,L)    \no \\
\tilde{\xi}_R(\tilde{U})  & \rightarrow & R \, \tilde{\xi}_R(\tilde{U}) \, 
             K^\dagger (\tilde{U},R,L)   .
\eeqa
The field $\tilde{U}$ from the last section is defined as
\beq
\tilde{U} = \tilde{\xi}_R \, \tilde{\xi}_L^\dagger .
\eeq
The baryon octet $B$ is given by the matrix
\beqa
 B =  \left(
\matrix { {1\over \sqrt 2} \Sigma^0 + {1 \over \sqrt 6} \Lambda
&\Sigma^+ &p \nonumber \\
\Sigma^-
        & -{1\over \sqrt 2} \Sigma^0 + {1 \over \sqrt 6} \Lambda & n
        \nonumber \\
\Xi^-
        &  \Xi^0 &- {2 \over \sqrt 6} \Lambda  \nonumber \\} 
\!\!\!\!\!\!\!\!\!\!\!\!\!\!\! \right) \, \, \, \, \,  
\end{eqnarray}
which transforms as a matter field
\beq
B \rightarrow B' = K B K^\dagger .
\eeq
The covariant derivative of the baryon fields reads
\beq
[ \tilde{D}_\mu , B] = \partial_\mu B + [ \tilde{\Gamma}_\mu , B]
\eeq
with $ \tilde{\Gamma}_\mu$ being the chiral connection
\beq
\tilde{\Gamma}_\mu = \frac{1}{2} \Big[ \tilde{\xi}_R^\dagger \, ( \partial_\mu 
    - i r_\mu ) \, \tilde{\xi}_R \, + \, \tilde{\xi}_L^\dagger \, 
       ( \partial_\mu - i l_\mu ) \, \tilde{\xi}_L  \Big] 
\eeq
and $r_\mu = v_\mu + a_\mu$, $l_\mu = v_\mu - a_\mu$.
For electromagnetic interactions the external fields are $a_\mu = 0$ and
$v_\mu = -e Q {\cal A}_\mu$ with the quark charge matrix $Q= \frac{1}{3}
\mbox{diag} (2,-1,-1)$.
In order to incorporate the interactions with the mesons 
into the effective theory it is convenient
to form an object of axial-vector type with one derivative
\beq
\tilde{\xi}_\mu = i \Big[ \tilde{\xi}_R^\dagger \, ( \partial_\mu 
    - i r_\mu ) \, \tilde{\xi}_R \, - \, \tilde{\xi}_L^\dagger \, 
       ( \partial_\mu - i l_\mu ) \, \tilde{\xi}_L  \Big] .
\eeq
Further ingredients of the non-linear representation are
\beq
\tilde{\chi}_{\pm} = \tilde{\xi}_L^\dagger \, \tilde{\chi}^\dagger 
   \, \tilde{\xi}_R \pm  \tilde{\xi}_R^\dagger \, \tilde{\chi} \, \tilde{\xi}_L
\eeq
and the quantity
\beq
\tilde{F}^{\pm}_{\mu \nu} = \tilde{\xi}_R^\dagger \, F^R_{\mu \nu} 
\, \tilde{\xi}_R \pm \tilde{\xi}_L^\dagger \, F^L_{\mu \nu} \, \tilde{\xi}_L ,
\eeq
where $F^{R/L}_{\mu \nu}$ are the field strength tensors of $r_\mu / l_\mu$.
The most general relativistic effective Lagrangian up to second order in the
derivative expansion and contributing to the electric dipole moments of the
neutron and $\Lambda$ reads
\beqa  \label{bar0}
{\cal L}_{\phi B} &=& i W_1 \langle [\tilde{D}^{\mu},\bar{B}]\gamma_{\mu} B
\rangle - i W_1^* \langle \bar{B}  \gamma_{\mu}  [\tilde{D}^{\mu},B] \rangle 
+ W_2 \langle \bar{B}B \rangle \no \\
&& + W_3 \langle \bar{B} \gamma_{\mu}
 \gamma_5 \{\tilde{\xi}^{\mu},B\} \rangle  
+  W_4 \langle \bar{B} \gamma_{\mu} \gamma_5 [\tilde{\xi}^{\mu},B] \rangle 
+ W_5 \langle \bar{B} \gamma_{\mu} \gamma_5 B \rangle \langle 
\tilde{\xi}^{\mu} \rangle  \no \\
&& + i W_6 \langle \bar{B} \gamma_5 B \rangle 
+ W_7 \langle \bar{B} \{ \tilde{\chi}_+, B \} \rangle 
+ W_8 \langle \bar{B} [ \tilde{\chi}_+, B ] \rangle  
+ W_9 \langle \bar{B} B \rangle \langle \tilde{\chi}_+ \rangle \no \\
&& + i W_{10} \langle \bar{B} \{ \tilde{\chi}_-, B \} \rangle 
+ i W_{11} \langle \bar{B} [ \tilde{\chi}_-, B ] \rangle  
+ i W_{12} \langle \bar{B} B \rangle \langle \tilde{\chi}_- \rangle \no \\
&& + i W_{13} \langle \bar{B} \sigma_{\mu \nu} \gamma_5 
      \{\tilde{F}_+^{\mu \nu}, B \} \rangle
+ i W_{14} \langle \bar{B} \sigma_{\mu \nu} \gamma_5 
      [\tilde{F}_+^{\mu \nu}, B ] \rangle\no \\
&& + i W_{15} \langle \bar{B} \sigma_{\mu \nu} \gamma_5 B \rangle
      \langle \tilde{F}_+^{\mu \nu} \rangle .
\eeqa
The $W_i$ are functions of the combination 
$\sqrt{6} \eta_0/F_0 +\bar{\theta}_0$.
From parity it follows that $W_{1,\ldots,5}$ and $W_{7,8,9}$ are even in  this
variable, whereas $W_6$ and $W_{10,\ldots,15}$ are odd.
The latter have not been taken into account in \cite{PR}.
One can further reduce the number of independent terms by making the
following transformation.
By decomposing the baryon fields into their left- and right handed 
components
\beq
B_{R/L} = \frac{1}{2} ( 1 \pm \gamma_5) B
\eeq
and transforming the left- and right-handed states separately via
\beqa  \label{trans}
B_{R/L} &\rightarrow& \frac{1}{\sqrt{W_2 \pm i W_6}} B_{R/L} \no \\
\bar{B}_{R/L} &\rightarrow& \frac{1}{\sqrt{W_2 \mp i W_6}} \bar{B}_{R/L} 
\eeqa
one can eliminate the $\langle \bar{B} \gamma_5 B \rangle $ term
and simplify the coefficient of $\langle \bar{B}B \rangle$. The details of this
calculation are given in \cite{B1}. 
Note that this transformation leads to mixing of the terms of the type 
$\bar{B} \tilde{\chi}_{\pm} B$ with $\bar{B} \gamma_5 \tilde{\chi}_{\pm} B$
which are of third chiral order and have been neglected
here. Furthermore, the terms $W_{13,\ldots,15}$ mix with the terms from the
Lagrangian
\beq  \label{mag}
{\cal L} = W_{16} \langle \bar{B} \sigma_{\mu \nu} 
      \{\tilde{F}_+^{\mu \nu}, B \} \rangle
+  W_{17} \langle \bar{B} \sigma_{\mu \nu} 
      [\tilde{F}_+^{\mu \nu}, B ] \rangle\no 
+  W_{18} \langle \bar{B} \sigma_{\mu \nu}  B \rangle
      \langle \tilde{F}_+^{\mu \nu} \rangle ,
\eeq
but in both cases the form of the Lagrangian does not change and we can proceed
by neglecting the $W_6$ term and setting $W_1 = W^*_1$ and $W_2 = - \mnod$ with
$\mnod$ being the baryon mass in the chiral limit \cite{B1}.
The expansion of the coefficients in terms of the $W_i$ read
\beqa
 && W_1 = - \frac{1}{2} + \ldots ,  \qq
    W_2 = - \mnod , \no \\
 && W_3 = - \frac{1}{2} D + \ldots , \qq
    W_4 = - \frac{1}{2} F + \ldots , \qq
    W_5 =  \frac{1}{2} \lambda + \ldots , \no \\
 && W_7 = b_D + \ldots , \qq
    W_8 = b_F + \ldots , \qq 
    W_9 = b_0 + \ldots ,  \no  \\
 && W_i = w_i \, 
     (\frac{\sqrt{6}}{F_0} \eta_0 +\bar{\theta}_0)  + \ldots ,  
     \qq \mbox{for} \q i = 10,\ldots ,15 \q ,
\eeqa
where the ellipses denote higher orders in 
$\sqrt{6} \eta_0/F_0 +\bar{\theta}_0$
and we have only shown terms that can contribute to one-loop order.
The axial-vector couplings $D$ and $F$ can be determined from semileptonic
hyperon decays. A fit to the experimental data delivers 
$D=0.80 \pm 0.01$ and $F=0.46 \pm 0.01$ \cite{CR}.
The third coupling, $\lambda$, is specific to the axial flavor-singlet baryonic
current.
The coefficients $b_{D,F,0}$ have been determined from the calculation of the
baryon masses and the $\pi N$ $\sigma$-term up to fourth chiral order
\cite{BM} and their mean values are, in units of GeV$^{-1}$,
\beq
b_D = 0.079, \qq    b_F = -0.316, \qq  b_0 = - 0.606 .
\eeq
The numerical values of the parameters $w_{10,\ldots,15}$ are not known.
We leave them undetermined for the time being and will give later an upper
bound.

Again the correct vacuum has to be chosen.
This is done by setting
\beqa
\tilde{\xi}_R  & = & \sqrt{U_0} \, \xi_R \no \\
\tilde{\xi}_L  & = & \sqrt{U_0^\dagger} \, \xi_L ,
\eeqa
and we will choose the coset representatives such that
\beq
\xi_R  = \xi_L^\dagger = u = \sqrt{U} .
\eeq
For diagonal external fields $v_\mu$ and $a_\mu$ one can write
\beq
\tilde{\Gamma}_\mu = \Gamma_\mu = 
    \frac{1}{2} \Big[ u^\dagger \, ( \partial_\mu 
    - i r_\mu ) \, u \, + \, u \, 
       ( \partial_\mu - i l_\mu ) \,  u^\dagger \Big] 
\eeq
and 
\beq
\tilde{\xi}_\mu = u_\mu = 
    i  \Big[ u^\dagger \, ( \partial_\mu 
    - i r_\mu ) \, u \, - \, u \, 
       ( \partial_\mu - i l_\mu ) \,  u^\dagger \Big] .
\eeq
Furthermore, one obtains
\beqa \label{chi}
\tilde{\chi}_+ & = & \chi_+ -i {\cal A} \, ( U-U^\dagger) 
                       -i {\cal B} \chi_- \no \\
\tilde{\chi}_- & = & \chi_- -i {\cal A} \, ( U+U^\dagger) 
                       -i {\cal B} \chi_+ ,
\eeqa
where the quark mass matrix enters in the combinations
\beq
\chi_\pm = u \chi^\dagger u \pm u^\dagger \chi u^\dagger .
\eeq
Finally, $\tilde{F}_{\mu \nu}^+$ simplifies to
\beq
\tilde{F}_{\mu \nu}^+ = F_{\mu \nu}^+ = u^\dagger F_{\mu \nu}^R u + 
    u F_{\mu \nu}^L u^\dagger .
\eeq
The relativistic baryon Lagrangian reads to the order we are working
\beqa  \label{bar1}
{\cal L}_{\phi B} &=&  i  \langle \bar{B}  \gamma_{\mu}  
[D^{\mu},B] \rangle - \mnod  \langle \bar{B}B \rangle \no \\
&& - \frac{1}{2} D \langle \bar{B} \gamma_{\mu}
 \gamma_5 \{u^{\mu},B\} \rangle  
-\frac{1}{2} F \langle \bar{B} \gamma_{\mu} 
  \gamma_5 [u^{\mu},B] \rangle 
+ \frac{1}{2} \lambda \langle \bar{B} \gamma_{\mu} \gamma_5 B \rangle \langle 
u^{\mu} \rangle  \no \\
&& - i b_D {\cal A} \langle \bar{B} \{ U - U^\dagger, B \} \rangle 
- i b_F {\cal A} \langle \bar{B} [ U - U^\dagger , B ] \rangle  
- i b_0 {\cal A} \langle \bar{B} B \rangle \langle U - U^\dagger \rangle \no \\
&& + 4  {\cal A} w_{10} \frac{\sqrt{6}}{F_0} \eta_0 
     \langle \bar{B} B  \rangle 
+ 6 {\cal A} w_{12} \frac{\sqrt{6}}{F_0} \eta_0
   \langle \bar{B} B \rangle \no \\
&& + i ( w_{13}' \bar{\theta}_0  + w_{13} \frac{\sqrt{6}}{F_0} \eta_0 )
    \langle \bar{B} \sigma_{\mu \nu} \gamma_5 
      \{F_+^{\mu \nu}, B \} \rangle \no \\
&& + i ( w_{14}' \bar{\theta}_0  + w_{14} \frac{\sqrt{6}}{F_0} \eta_0 )
    \langle \bar{B} \sigma_{\mu \nu} \gamma_5 
      [F_+^{\mu \nu}, B ] \rangle\no \\
&& + i( w_{15}' \bar{\theta}_0  + w_{15} \frac{\sqrt{6}}{F_0} \eta_0 )
     \langle \bar{B} \sigma_{\mu \nu} \gamma_5 B \rangle
      \langle F_+^{\mu \nu} \rangle ,
\eeqa
where we have neglected meson-baryon interactions with more than one meson
field since they do not contribute at one-loop order
and terms of ${\cal O}(\bar{\theta}_0^2)$ or higher orders are omitted
throughout this work.
Note that there exist terms of fourth chiral order of the type $ \bar{B}
\sigma_{\mu \nu} \gamma_5 \tilde{\chi}_- \tilde{F}_+^{\mu \nu} B$. Using
Eq. (\ref{chi}) for $ \tilde{\chi}_-$ they induce $CP$-violating terms of the
form $ \bar{\theta}_0 \bar{B}
\sigma_{\mu \nu} \gamma_5  \tilde{F}_+^{\mu \nu} B$
which are already accounted for by the terms $w_{13,14,15}$.
This amounts to a renormalization of the couplings 
$\bar{\theta}_0 w_{13,14,15}$.
We have therefore introduced the notation $w_{13,14,15}'$ for these
interaction terms, in order to distinguish them from the unrenormalized
$w_{13,14,15}$ of the interactions proportional to $\eta_0$.

The drawback of the relativistic framework including baryons is that 
due to the existence of a new mass scale, namely the baryon mass
in the chiral limit $\mnod$, there exists no strict
chiral counting scheme, i.e. a one-to-one correspondence between the meson
loops and the chiral expansion. In order to overcome this problem
one integrates out the heavy degrees of freedom of the baryons, similar to a
Foldy-Wouthuysen transformation, so that a chiral counting scheme emerges.
To this end, one constructs eigenstates of the velocity projection
operator $P_v =  ( 1 + v\! \! /)/2$
\beq
B_v (x) = e^{i \mnod \, v \cdot x} \: P_v \, B (x) .
\eeq
The Dirac algebra simplifies considerably. It allows to express any Dirac 
bilinear $\bar{B_v} \Gamma_{\mu} B_v \, (\Gamma_{\mu} = 1, \gamma_{\mu}, 
\gamma_5, \ldots ) $ in terms of the velocity $v_{\mu}$ and the spin operator
$ 2 S_{\mu} = i \gamma_5 \sigma_{\mu \nu} v^{\nu} $. 
One can rewrite the Dirac bilinears which appear in the present calculation 
as 
\beqa
\bar{B_v} \gamma_\mu \gamma_5 B_v   =   2 \bar{B_v} S_{\mu} B_v  , \qq  
\bar{B_v} \sigma_{\mu \nu} \gamma_5  B_v   =   2 i \, (
   v_{\mu} \bar{B_v} S_\nu   B_v  -  v_{\nu} \bar{B_v} S_\mu   B_v \, )   .
\eeqa
In the following, we will drop the index $v$.
The Lagrangian of the heavy baryon formulation reads
\beqa  \label{bar2}
{\cal L}_{\phi B} &=&  i  \langle \bar{B}   [v \cdot D,B] \rangle 
-  D \langle \bar{B} S_{\mu} \{u^{\mu},B\} \rangle  
 - F \langle \bar{B} S_{\mu} [u^{\mu},B] \rangle 
+ \lambda \langle \bar{B} S_{\mu} B \rangle \langle 
u^{\mu} \rangle  \no \\
&& - i b_D {\cal A} \langle \bar{B} \{ U - U^\dagger, B \} \rangle 
- i b_F {\cal A} \langle \bar{B} [ U - U^\dagger , B ] \rangle  
- i b_0 {\cal A} \langle \bar{B} B \rangle \langle U - U^\dagger \rangle \no \\
&& + 4 {\cal A} w_{10} \frac{\sqrt{6}}{F_0} \eta_0 
     \langle \bar{B} B  \rangle 
+ 6 {\cal A} w_{12} \frac{\sqrt{6}}{F_0} \eta_0
   \langle \bar{B} B \rangle \no \\
&& - 4 ( w_{13}' \bar{\theta}_0 + w_{13} \frac{\sqrt{6}}{F_0} \eta_0 )
    \langle \bar{B} v_\mu S_\nu 
      \{F_+^{\mu \nu}, B \} \rangle \no \\
&& -4  ( w_{14}' \bar{\theta}_0  + w_{14} \frac{\sqrt{6}}{F_0} \eta_0 )
    \langle \bar{B}  v_\mu S_\nu
      [F_+^{\mu \nu}, B ] \rangle\no \\
&& - 4 ( w_{15}' \bar{\theta}_0  + w_{15} \frac{\sqrt{6}}{F_0} \eta_0 )
     \langle \bar{B} v_\mu S_\nu  B \rangle
      \langle F_+^{\mu \nu} \rangle .
\eeqa
No relativistic corrections are needed to the order we are working.

\section{The electric dipole moments $d_n^\gamma$ and $d_\Lambda^\gamma$}
In this section, we will calculate the electric dipole moments of the neutron
and the $\Lambda$ at lowest order in chiral perturbation theory, i.e. 
${\cal O}(p^2)$. Due to the vacuum alignment the baryon Lagrangian contains
interaction terms $b_{D, F, 0}$ of zeroth chiral order and one-loop
diagrams with these vertices will also contribute at ${\cal O}(p^2)$.
In the relativistic framework of \cite{PR} the electric dipole moment of 
the neutron $d_n^\gamma$ has been defined via
\beq
{\cal L}_{nEDM} = \frac{1}{2} d_n^\gamma \bar{n} i \sigma_{\mu \nu} \gamma_5 
n  F^{\mu \nu} ,
\eeq
where $F^{\mu \nu}$ is the field strength tensor of the photon field $A^\mu$.
We prefer to rewrite this as a form factor
\beq
D_n^\gamma(q^2) \bar{u}(p') \sigma_{\mu \nu} \gamma_5 u(p) q^\mu
\eeq
with $q= p' -p$ being the momentum transfer.
The electric dipole moment is given by
\beq
d_n^\gamma = D_n^\gamma(q^2= 0).
\eeq
For the calculation of the form factor in the heavy baryon approach we set
$v_\mu =(1,{\bf 0})$ and use the Breit frame $v \cdot p = v \cdot p'$ since it
allows a unique translation of Lorentz-covariant matrix elements into
non-relativistic ones.
In this frame the form factor reads
\beq
- 2 i D_n^\gamma(q^2) \bar{H} v_{\nu} S \cdot q  H + \ldots ,
\eeq
where $H$ is the large component of $u$ and the ellipsis stands for a similar
expression in the small components of $u$, which is of higher chiral order and
will be omitted.
The electric dipole moment 
$d_n^\gamma$ receives contributions from the $w_{13}$-term and
the loops (we work in the isospin limit $m_u=m_d=\hat{m}$)
\beq
d_n^\gamma = d_n^{\gamma \, (tree)} + d_n^{\gamma \, (loop)}
\eeq
with
\beq  \label{wr}
d_n^{\gamma \, (tree)} = - 8 \, e \, \bar{\theta}_0 
\Big[ \, \frac{1}{3} w_{13}^r
+ \frac{16}{F_\pi^2 F_0^2 m_{\eta_0}^2} V_3^{(1)} V_0^{(2)}  w_{13} \, \Big] ,
\eeq
where the pertinent diagrams are shown in Figure 1.
Diagram 1b) is missing in \cite{PR} since the mesonic Lagrangian used within
this work does not have a term linear in the singlet field $\eta_0$.
Such an interaction originates from the term $V_3 \langle \chi_- \rangle$ which
is not considered in \cite{PR}.
The chiral logarithms of the diagrams shown in Fig. 2 read
\beq  \label{lop}
d_n^{\gamma \, (loop)} = \frac{1}{\pi^2 F_\pi^4} e V_0^{(2)} \, \bar{\theta}_0
 \Big[ - (b_D + b_F) ( D+F) \ln \frac{m_\pi^2}{\mu^2}
+ (b_D - b_F) ( D-F) \ln \frac{m_K^2}{\mu^2} \Big]
\eeq
with $\mu$ being the scale introduced in dimensional regularization.
In the Breit frame only diagrams 2a) and b) contribute to the electric dipole
moment. Diagrams 2c), d) vanish and 2e), f) are proportional to $S_\nu$
and therefore do not contribute to $d_n^\gamma$.
The loop integral 
for diagrams 2a) and b) contains also analytic and divergent pieces
which can be absorbed by redefining $w_{13}'$. The divergent pieces of
$w_{13}'$ cancel the divergencies from the loops and render the final
expression finite. We summarize the remaining analytical contributions in
$w_{13}^r$, so that $w_{13}^r$ in
Eq. (\ref{wr}) is understood to be finite.
The results  in Eqs. (\ref{wr}) and (\ref{lop}) are not in contradiction with
the fact that the electric dipole moment induced by the $\theta_0 $-term
tends to
zero if any of the quark masses vanish. If, e.g., $m_u =0$ then a solution for 
Eq. (\ref{cof}) is given by $\varphi_u = \theta_0$ and $\varphi_{d,s}=0$
leading to $\bar{\theta}_0 = \theta_0 - \varphi_u =0$. Therefore,
$d_n^{\gamma}$ vanishes in this case.

The results for the $\Lambda$ are
\beq  
d_\Lambda^{\gamma \, (tree)} = - 4 \, e \, \bar{\theta}_0 
\Big[ \, \frac{1}{3} w_{13}^r
+ \frac{16}{F_\pi^2 F_0^2 m_{\eta_0}^2} V_3^{(1)} V_0^{(2)}  w_{13} \, \Big] 
\eeq
and
\beq  \label{lambda}
d_\Lambda^{\gamma \, (loop)} = - \frac{1}{\pi^2 F_\pi^4} e V_0^{(2)} \, 
\bar{\theta}_0 \, [b_D F + b_F D] \, \ln \frac{m_K^2}{\mu^2}  .
\eeq
Note that the relation $ d_n^\gamma = 2 d_\Lambda^\gamma $ is only valid at
tree level and not for the chiral logarithms as claimed in \cite{PR}.
The discrepancy is due to the lack of pion loops in the present work. Once one
accounts for the mistake made in \cite{PR} by replacing $\ln m_\pi$ by
$\ln m_K$ in the chiral loop contribution for $d_\Lambda^\gamma$ one obtains
our result (\ref{lambda}).

\section{Numerical results and conclusions}
In order to compute the numerical results for the electric dipole moments shown
in the last section, we use the central values for the parameters
$b_D = 0.079$ GeV$^{-1}$, $b_F = -0.316$ GeV$^{-1}$, $D= 0.80$ and $F= 0.46$.
To lowest order in the angles $\varphi_q$ and using $\hat{m} \ll m_s$ one can
express $\bar{\theta}_0$ in terms of $\theta_0$ via
\beq
\theta_0 \simeq [ 1 + \frac{8 V_0^{(2)}}{F_\pi^2 m_\pi^2}]  \bar{\theta}_0.
\eeq
From the calculation of the $\eta$ and $\eta'$ masses and decay constants 
one can extract the value for $V_0^{(2)}$  \cite{H-S2}
\beq
V_0^{(2)} \simeq - \frac{27}{4} F_\pi^4 
\simeq  - 5.0 \times 10^{-4}\; \mbox{GeV}^{4} ,
\eeq
so that
\beq
\bar{\theta}_0 \simeq \frac{F_\pi^2 m_\pi^2}{8 V_0^{(2)}} \, \theta_0 
\simeq -0.04 \,  \theta_0 .
\eeq
Inserting this into the loop contribution and using $\mu = 1$ GeV, $m_{\eta_0}
\simeq  m_{\eta'} = 958$ MeV we obtain
\beqa
d_n^{\gamma \, (loop)}  & = &  - 7.5 \times 10^{-16} \,  \theta_0
\, e \, \mbox{cm}   \no \\
d_\Lambda^{\gamma \, (loop)}   & = &   - 1.7 \times 10^{-16} \,  \theta_0 \,
   e \, \mbox{cm} .
\eeqa
The numerical result for $d_n^{\gamma \, (loop)}$ is in agreement with the one
given in \cite{PR} once one accounts for the different values of the parameters
$b_D, b_F, D, F$ and the scale $\mu$ used within that work.
The result for the chiral logarithm of $d_\Lambda^{\gamma \, (loop)}$ is
considerably smaller than for $d_n^{\gamma \, (loop)}$ since there is no
contribution from the pion loops which dominate in the case of $d_n^\gamma$.

A precise numerical value for the tree contribution to the electric dipole
moments cannot be given since the parameters $w_{13}$ and $w_{13}^r$
are not known. However, we will give an upper bound for their contribution
based on large $N_c$ arguments which seem to work well in the purely mesonic
sector \cite{H-S2, KL}.
In the present investigation we are only interested in an order of magnitude
estimate for the vacuum angle $\theta$ and for this purpose it is sufficient to
give a numerical range for the tree level contribution.
We will first estimate the ratio of diagrams 1a) and 1b) using 1/$N_c$
arguments. Applying large  $N_c$ counting rules, see e.g. \cite{Leu, H-S},
both $w_{13}$ and $w_{13}^r$ are of order ${\cal O}(N_c^0)$ so that we can
assume  $|w_{13}/w_{13}^r| = {\cal O}(1)$.
One obtains the ratio
\beq
\frac{|d_n^{\gamma \, (1b)}|}{|d_n^{\gamma \, (1a)}|}
= \frac{|d_\Lambda^{\gamma \, (1b)}|}{|d_\Lambda^{\gamma \, (1a)}|}
\simeq \frac{48}{F_\pi^4 m_{\eta'}^2} |V_0^{(2)} V_3^{(1)}|
\simeq  0.12 ,
\eeq
where we have used $F_\pi/F_0 =  1 + {\cal O}(N_c^{-1})$  
and taken the value for $V_3^{(1)}$ from \cite{H-S2}
\beq
V_3^{(1)} \simeq  0.04 F_\pi^2 \simeq  3.5 \times 10^{-4}\; \mbox{GeV}^{2} .
\eeq
Diagram 1b) turns out to be insignificant in our estimate.

Based on large $N_c$ arguments one can also give an upper bound for $w_{13}^r$.
The contact terms of the Lagrangian in Eq. (\ref{mag}), which contribute to the
magnetic moments of the baryons, describe -- similar to the $w_{13}^r$-term --
the coupling of the field strength tensor $F_{\mu \nu}^+$ to the
baryons. 
But the leading coefficient of the Taylor expansion of $W_{16}$, $w_{16}$, 
is of order ${\cal O}(N_c^1)$, whereas $|w_{13}^r| \simeq |w_{13}| = 
{\cal O}(N_c^0)$, so that we can assume $|w_{13}^{(r)}| < |w_{16}|$.
In a calculation of the baryon magnetic moments to fourth chiral order 
$w_{16}$ has been determined to be  \cite{SM}
\beq
w_{16} \simeq 0.4 \; \mbox{GeV}^{-1}  .
\eeq
Inserting this upper limit for $w_{13}^r$ one obtains
\beqa  
|d_n^{\gamma \, (tree)}| & < &  9.6 \times 10^{-16} \,  \theta_0
\,  e \, \mbox{cm} \no \\
|d_\Lambda^{\gamma \, (tree)}| & < &  4.8 \times 10^{-16} 
  \,  \theta_0 \,  e \, \mbox{cm} .
\eeqa
Since we have taken quite conservative limits, 
the tree level contributions could
be dominating if the extreme values are chosen. A more realistic
estimate might be obtained by setting 
$ |w_{13}^{(r)}/ w_{16}| \simeq \frac{1}{3}$ for $N_c=3$, so that
\beqa   \label{bound}
|d_n^{\gamma \, (tree)}| & \simeq &  3.2 \times 10^{-16} \,  \theta_0
 \,  e \, \mbox{cm} \no \\
|d_\Lambda^{\gamma \, (tree)}| & \simeq &  1.6 \times 10^{-16} \,  \theta_0
\,  e \, \mbox{cm} .
\eeqa
While the chiral logarithms dominate for $d_n^\gamma$, the tree level
contribution can be substantial for the $\Lambda$.
This leads to
\beqa
d_n^\gamma =  (-7.5 \pm  3.2)  \times 10^{-16}  \,  \theta_0
\,   e \, \mbox{cm}  \no \\
d_\Lambda^\gamma =  (-1.7 \pm  1.6)  \times 10^{-16}  \,  \theta_0
\,   e \, \mbox{cm}  ,
\eeqa
where the theoretical uncertainty is given by the estimate of the tree
level contribution in Eq. (\ref{bound}).
From a comparison of the central value for $d_n^\gamma$ with the experimental 
upper limit in Eq. (\ref{exp})
we obtain
\beq
|\theta_0| <  8.4  \times 10^{-11}   .
\eeq
In the case of the $\Lambda$ the experimental constraint is given by
\beq
d_\Lambda^\gamma <  1.5  \times 10^{-16}   \, e \,\mbox{cm}
\eeq
which leads to
\beq
|\theta_0| <  0.9   .
\eeq
We have shown that it is possible to obtain a reliable limit for the vacuum 
angle $\theta_0$ by calculating the electric dipole moment of the neutron
within the framework of heavy baryon chiral perturbation theory.
To this end, we have constructed the most general effective Lagrangian up to
one-loop order in the presence of the vacuum angle $\theta_0$ with the
method proposed in \cite{B1}.
The theoretical uncertainty from unknown parameters at tree level has been
estimated by using large $N_c$ arguments. While the chiral loops dominate for
the electric dipole moment of the neutron, the counterterms can be substantial
in the case of the $\Lambda$.

\section*{Acknowledgments}
The author thanks 
N. Kaiser for helpful discussions and careful reading of the manuscript.

\newpage


\section*{Figure captions}

\begin{enumerate}

\item[Fig.1] Shown are the tree diagrams for the electric dipole moment.
         Solid and dashed lines denote baryons and
         pseudoscalar mesons, respectively.  The wavy line represents
         a photon and the dot is a
         $CP$-violating vertex.

\item[Fig.2] Loop diagrams contributing to the electric dipole moment. 
         Solid and dashed lines denote baryons and
         pseudoscalar mesons, respectively.  The wavy line represents
         a photon and the dot is a
         $CP$-violating vertex.

\end{enumerate}

\newpage

\begin{center}
 
\begin{figure}[bth]
\centering
\centerline{
\epsfbox{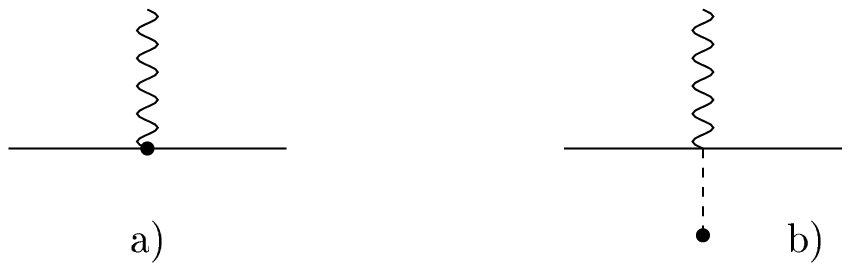}}
\end{figure}

\vskip 0.7cm

Figure 1

\vskip 1.5cm

\begin{figure}[tbh]
\centering
\centerline{
\epsfbox{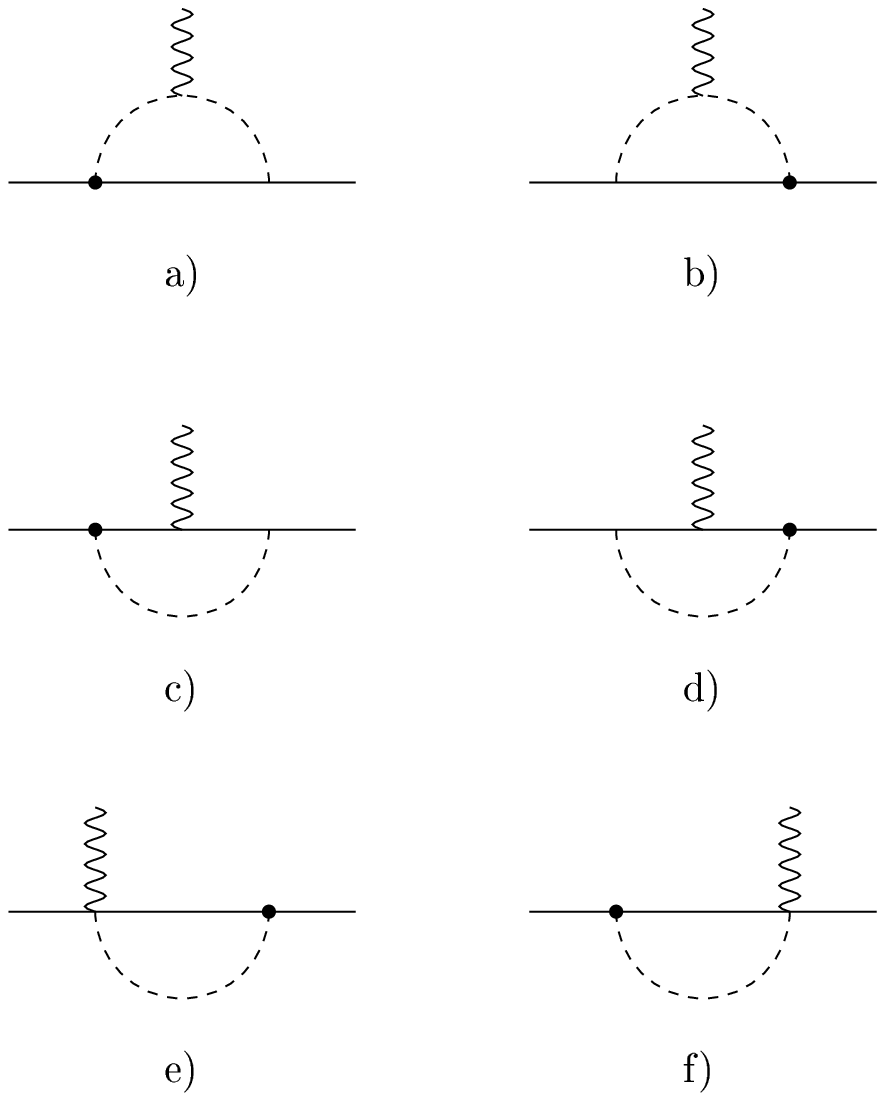}}
\end{figure}

\vskip 0.7cm

Figure 2

\end{center}

\end{document}